\documentclass[aps,prl,groupedaddress,showpacs,preprint]{revtex4}
\usepackage{graphicx}
\usepackage[dvips]{epsfig}
\usepackage{dcolumn}
\usepackage{subfigure}%
\usepackage{bm}
\usepackage{amssymb}
\usepackage{amsmath}

\begin{document}

\title{Nonlinear field theories during homogeneous spatial dilation}

\author{Carlos Escudero}

\affiliation{Departamento de Matem\'aticas \& ICMAT (CSIC-UAM-UC3M-UCM), Universidad Aut\'onoma de Madrid, E-28049 Madrid, Spain}

\begin{abstract}
The effect of a uniform dilation of space on stochastically driven nonlinear field theories is examined.
This theoretical question serves as a model problem for examining the properties of nonlinear field theories
embedded in expanding Euclidean Friedmann-Lema\^{\i}tre-Robertson-Walker metrics in the context of
cosmology, as well as different systems in the disciplines of statistical mechanics and
condensed matter physics. Field theories are
characterized by the speed at which they propagate correlations within themselves. We show that for linear field
theories correlations stop propagating if and only if the speed at which the space dilates is higher than
the speed at which correlations propagate. The situation is in general different for nonlinear field theories.
In this case correlations might stop propagating even if the velocity at which space dilates is lower than
the velocity at which correlations propagate. In particular, these results imply that it is not possible to
characterize the dynamics of a nonlinear field theory during homogeneous spatial dilation {\it a priori}.
We illustrate our findings with the nonlinear Kardar-Parisi-Zhang equation.
\end{abstract}


\maketitle

Field theories are one of the most powerful theoretical tools in the physics of spatially extended systems. Understanding their mathematical structure
is concomitant to getting a deeper understanding of the structure and behavior of these systems, and even to the discovery of new physics.
In this work, we will consider stochastic field theories immersed in spatial domains whose size grows in time. As will be shown herein, this effect
can have nontrivial consequences on the dynamics of such fields. This phenomenology is present, for example, in cosmology, whenever it is necessary to
consider an universe which is expanding in time. One of these instances is inflationary cosmology that studies the exponentially fast growth
of the universe at its early stages. In fact, the importance of inflation is very large as it is assumed to be the responsible for the current
homogeneous and isotropic appearance of the universe. This phenomenology has also a very natural statistical mechanical motivation.
First of all, the models we are going to consider are paradigmatic in the theory of dynamic critical phenomena~\cite{ks3,ks4}.
In this context one would like to determine whether or not fluctuations are able to break the homogeneity of a space which
undergoes a uniform dilation, given that this mechanism acts as a homogenization on the large scale. Biological systems have a connection with it too,
mainly in the context of pattern formation and generation of form during growth~\cite{arbor,escuderocsf}. Indeed, the equations this work is focused on can
be considered as hydrodynamic descriptions of the model Eden introduced as an idealized description of a growing cell colony~\cite{eden1,eden2}.

We will focus on stochastic field theories whose correlation length can be explicitly computed.
When these theories are embedded in a spatial domain which is uniformly expanding in time, what
mathematically reduces to considering a Friedmann-Lema\^{\i}tre-Robertson-Walker (FLRW) metric,
an effective loss of correlation takes place whenever the expansion
is fast enough. Intuitively, determining the threshold of the speed of expansion that causes this loss of correlation would mean finding the condition that
ensures the macroscopic appearance of the corresponding system is homogeneous and isotropic.
Although in reality this condition is just a necessary but not sufficient one,
it is still one of the key ingredients in the search for determining the macroscopic appearance of one such system.

In this work we are concerned with stochastic fields $\phi(x,t)$ for which the spatial coordinate $x \in \mathbb{R}^d$ and the temporal coordinate
$t \in \mathbb{R}^+$. This field will obey the
generic equation of motion
\begin{equation}
\partial_t \phi = f(\hat{L}_1 \phi, \hat{L}_2 \phi, \cdots) + \chi(x,t),
\end{equation}
where the $\hat{L}_i$'s, $i=1,2,\cdots$, are linear differential or integro-differential operators acting on the field, $f$ is an in general nonlinear function
of its arguments and $\chi$ denotes a space-time noise to be specified in the following.

Field theories can be characterized by sets of exponents. One of them is the dynamic exponent $z$ which characterizes the velocity at which
correlations propagate. If $\ell$ is the correlation length of one such theory then $\ell(t) \sim t^{1/z}$.

We will consider spatial dilation as a simple transformation of the spatial coordinates $x \to b(t) x$, where $b(t)$ is independent of $x$, $b(t)>1$ for $t>t_0$ and $b(t_0)=1$, and $t_0>0$ is the absolute origin of time. So this transformation is a strict dilation.

Two different correlation functions will be examined: the two-point function and the field difference correlation function.
For the models under consideration both will undergo dynamic scale invariance; so explicitly they read, the first one
\begin{equation}\label{corr2}
G_2(x,x';t)^2:=\left\langle \phi(x,t) \phi(x',t) \right\rangle = |x-x'|^{2\alpha}\mathcal{F} \left( \frac{|x-x'|}{t^{1/z}} \right),
\end{equation}
and the second one
\begin{equation}\label{corrd}
G_d(x,x';t)^2:=\left\langle |\phi(x,t) - \phi(x',t)|^2 \right\rangle = |x-x'|^{2\alpha}\mathcal{G} \left( \frac{|x-x'|}{t^{1/z}} \right),
\end{equation}
where role of the dynamic exponent is evident and the $\alpha$ exponent describes the variation of the field on a determined length scale~\cite{foot1};
$\mathcal{F}$, $\mathcal{G}$ are the scaling functions.

We say that a \emph{linear superposition principle} holds whenever, upon applying the dilation transformation $\{x,x'\} \to b(t) \{x,x'\}$, the correlation functions read
\begin{equation}\label{superprin}
G_{\{2,d\}}^2=b(t)^{2\alpha}|x-x'|^{2\alpha} \{\mathcal{F},\mathcal{G}\} \left( \frac{b(t)|x-x'|}{t^{1/z}} \right),
\end{equation}
for the \emph{same} exponents $\alpha$ and $z$.
It is clear where this expression comes from: in this case the internal dynamics of the field and the spatial dilation effect are simply superposed.

We start our discussion with the following family of linear equations:
\begin{equation}\label{linfield}
\partial_t \phi = - \nu |\nabla|^\zeta \phi + \xi(x,t),
\end{equation}
where the noise is assumed to be Gaussian with zero mean and correlation
$\left\langle \xi(x,t) \xi(x',t') \right\rangle= D \delta(x-x') \delta(t-t')$,
and the operator $|\nabla|^\zeta$ (we will always consider $\zeta>0$) is to be interpreted in the Fourier transform sense
$(|\nabla|^\zeta \phi)\hat{}=|k|^\zeta \hat{\phi}$. This operator accounts for the anomalous diffusion of the field, and its effect on this
type of theories has already been considered, even in the nonlinear case~\cite{katzav1}.
This model can be exactly integrated and shown to obey the above mentioned scalings with $z=\zeta$.
Actually, the presence of noise in this equation is trivial in the sense that the exponent $z$ does not change if
we set $D=0$.

Now we will apply the dilation transformation to Eq.~(\ref{linfield}),
and for the sake of concreteness we set $b(t)=(t/t_0)^\gamma$, where the \emph{growth index} $\gamma>0$.
One question naturally arises: is there linear superposition?
Even in this case in which the equation is linear the answer is only partially positive (and so in general it is negative). Linear superposition
holds for both correlation functions, without amendments, only if $\gamma < 1/\zeta$. For $\gamma > 1/\zeta$ the two-point function does not adopt the form described by
Eq.~(\ref{superprin})~\cite{escuderocsf,escudero0,escudero,escudero2}.
Indeed, this value of $\gamma$ plays a special role. For $\gamma<1/\zeta$ one can read from Eq.~(\ref{superprin}) that correlations
still propagate as time evolves. The contrary would happen if we reversed the inequality.
Although linear superposition does not take place for the two point function and large enough
$\gamma$, we still have a weaker yet intuitive result: propagation of correlations stops whenever $\gamma > 1/\zeta$~\cite{escudero,escudero2,escuderoap}.

Explicitly, after applying the dilation transformation, the equation of motion reads
\begin{equation}
\partial_t \phi = - \left( \frac{t_0}{t} \right)^{\zeta \gamma} \nu |\nabla|^\zeta \phi + \left( \frac{t_0}{t} \right)^{d \gamma/2}
\xi(x,t).
\end{equation}
Following Eq.~(\ref{superprin}) and for $\gamma < 1/\zeta$ one may define an effective dynamic exponent $z_{\mathrm{eff}} = \zeta/(1-\gamma
\zeta)$~\cite{escudero}. Thus in the limit $\gamma
\to 0^+$ one recovers the classical case $z_{\mathrm{eff}} \to
\zeta$, and when $\gamma \to (1/\zeta)^-$ then $z_{\mathrm{eff}} \to
\infty$. So we may talk about the decorrelation threshold
$\gamma_d = 1/\zeta$.

The main conclusion of this analysis is that for a rather general family of linear equations the decorrelation threshold is the intuitive one. We will show
that for nonlinear equations things are in general different. To this end one needs to introduce a nonlinear field theory whose dynamic exponent is nontrivial.
One such theory is given by the Kardar-Parisi-Zhang (KPZ) equation~\cite{kpz}
\begin{equation}\label{kpz}
\partial_t \phi= \nu \nabla^2 \phi + \frac{\lambda}{2} (\nabla \phi)^2 + \xi(x,t).
\end{equation}
Together with the interest of this equation in the fields of condensed matter and statistical physics~\cite{wioetal}, one finds its relevance in
cosmology~\cite{berera,barbero,dominguez,mercader,jones,coles,mamo,chavani,gusash}.
Two of the reasons underlying this universal character are the connection of Eq.~(\ref{kpz}) with the Burgers equation through the definition of the velocity
field ${\bf v}:= \nabla \phi$ and with the imaginary time Schr\"{o}dinger equation with a random potential by means
of the change of variables $\psi:=\exp[\lambda \phi/(2 \nu)]$.
If we set $D=0$ in this equation we find that $z=2$, as can be read from its exact solution~\cite{kpz}. However,
once the noise is switched on, the dynamic exponent becomes a function of the spatial dimension, $z=z(d)$. In particular
$z(1)=3/2$ and $z(2) \approx 1.7 <2$. We note that while the one-dimensional result is exact, the two-dimensional one is usually obtained numerically.
This is so because the calculation of this value has escaped all sorts of analytical approaches, with the notable exception of the method known as the Self-Consistent Expansion (SCE)~\cite{se1,se2}. Due to the remarkable success of this scheme
in finding the scaling behavior of this as well as different models~\cite{katzav2,katzav3,katzav4} we will rely on its results in the following.
Another value we will also be using is $z(4) \approx 1.8$.

Our aim is calculating the decorrelation threshold for the KPZ
equation. Following the linear theory one could na\"{\i}fly
expect $\gamma_d = 1/z$, and in particular $\gamma_d=2/3$ in $d=1$. In fact, the one dimensional result is presumably correct. Simulations of a discrete model
in the KPZ universality class have corroborated so~\cite{pastor}. Things are however different in higher dimensions as we will subsequently show.

Now we apply to Eq.~(\ref{kpz}) the dilation transformation
$x \longrightarrow \left( t/t_0 \right)^\gamma x$:
\begin{equation}
\label{kpzgdomain}
\partial_t \phi= \nu \left( \frac{t_0}{t} \right)^{2\gamma} \nabla^2 \phi + \frac{\lambda}{2} \left( \frac{t_0}{t} \right)^{2\gamma} (\nabla \phi)^2
+ \left( \frac{t_0}{t} \right)^{d\gamma/2}
\xi(x,t).
\end{equation}
So we will study this equation which
describes KPZ dynamics in an environment which is
\emph{undergoing spatial dilation} as time evolves.
This is a Langevin equation whose associated Fokker-Planck equation reads
\begin{equation}
\partial_t \mathcal{P} = \left( \frac{t_0}{t} \right)^{2 \gamma} \int
dx \frac{\delta}{\delta \phi} \left[ \nu \nabla^2 \phi +
\frac{\lambda}{2} (\nabla \phi)^2 \right] \mathcal{P} + \frac{D}{2}
\left( \frac{t_0}{t} \right)^{d \gamma} \int dx
\frac{\delta^2}{\delta \phi^2} \mathcal{P},
\end{equation}
where the solution $\mathcal{P}$ is the functional probability
distribution. This equation can be transformed to
\begin{equation}
\frac{\partial \mathcal{P}}{\left( \frac{t_0}{t} \right)^{2
\gamma}\partial t} = \int dx \frac{\delta}{\delta \phi} \left[ \nu
\nabla^2 \phi + \frac{\lambda}{2} (\nabla \phi)^2 \right] \mathcal{P} +
\frac{D}{2} \left( \frac{t_0}{t} \right)^{(d-2) \gamma} \int dx
\frac{\delta^2}{\delta \phi^2} \mathcal{P}.
\end{equation}
Now we change the temporal variable
\begin{equation}\label{change}
d\tau = \left( \frac{t_0}{t} \right)^{2 \gamma}  dt
\longrightarrow \tau = \frac{t_0^{2 \gamma}}{1-2\gamma} (t^{1-2
\gamma} - t_0^{1-2 \gamma}).
\end{equation}
We start assuming $\gamma < 1/2$ so
$\tau \approx (t_0^{2 \gamma})/(1-2\gamma) t^{1-2 \gamma}$ when $t \to \infty$.
After performing the change of variables and going back to the
Langevin equation we find
\begin{equation}
\partial_\tau \phi = \nu \nabla^2 \phi + \frac{\lambda}{2} (\nabla \phi)^2
+ \left( \frac{1-2\gamma}{t_0} \right)^{\frac{(2-d)\gamma}{2(1-
2\gamma)}} \tau^{\frac{(2-d)\gamma}{2(1- 2\gamma)}} \,
\xi(x,\tau).
\end{equation}
This noise rends in general a more difficult analytical
treatment due to the explicit $\tau-$dependence of its amplitude.
However, the situation becomes considerably simpler in $d=2$. In this case we recover
the KPZ equation for time $\tau$, so the dynamic exponent for this time variable
is $z'= z_{\mathrm{KPZ}}(d=2)$. Thus the
effective dynamic exponent for actual time $t$ is
$z_\mathrm{eff} = z_{\mathrm{KPZ}}(d=2)/(1-2\gamma)$.
Consequently decorrelation appears in the limit $\gamma \to (1/2)^- \Rightarrow
\gamma_d=1/2$, instead of $\gamma_d = 1/z_{\mathrm{KPZ}}(d=2) \approx 0.6
>1/2$. So the decorrelation threshold is anticipated, and this
counterintuitive result implies that a simple superposition
principle does not hold in this case.

Although this result proves the decorrelation threshold by itself, it is easy to compute the exact behavior at the value $\gamma=1/2$.
In this case one can correspondingly modify change of variables~(\ref{change}) to find $\tau=t_0 \ln (t/t_0)$. So for the critical value
of $\gamma$ correlations propagate logarithmically slow, and we find the effective value $z_\mathrm{eff}=\infty$.
For $\gamma >1/2$ change of variables~(\ref{change})
is still valid. One again finds that the solution of Eq.~(\ref{kpzgdomain}) becomes the solution of the classical KPZ equation in time $\tau$.
The particularity of this situation is that, as time $t$ progresses, time $\tau$ evolves from $\tau=0$ when $t=t_0$ to the finite value $\tau=t_0/(2\gamma-1)$
in the limit $t \to \infty$. So the resulting profile of the solution to Eq.~(\ref{kpzgdomain}) becomes the profile of the solution to Eq.~(\ref{kpz})
quenched at time $t_0/(2\gamma-1)$ asymptotically in time.

This effect is not purely two-dimensional. We now move to higher dimensions and
consider again the KPZ equation but with a different stochastic forcing
\begin{equation}\label{kpz2}
\partial_t \phi= \nu \nabla^2 \phi + \frac{\lambda}{2} (\nabla \phi)^2 + \eta(x,t),
\end{equation}
where the noise is Gaussian, has zero mean and its correlation reads
$\left\langle \eta(x,t) \eta(x',t') \right\rangle= D |x-x'|^{2\rho-d} \delta(t-t')$,
where $\rho>0$ specifies the degree of spatial correlation ($\rho=0$ sends us back to the spatially uncorrelated noise).
The equation resulting from applying the spatial dilation transformation to this one can again be mapped onto a Fokker-Planck description.
And again, the same transformation
$\tau = t_0^{2 \gamma} (t^{1-2 \gamma} - t_0^{1-2 \gamma})/(1-2\gamma)$,
yields in the limit $t \to \infty$ the equation
\begin{equation}
\partial_\tau \phi= \nu \nabla^2 \phi + \frac{\lambda}{2} (\nabla \phi)^2 + \left( \frac{1-2\gamma}{t_0} \right)^{\frac{(2+2\rho-d)\gamma}{2(1-
2\gamma)}} \tau^{\frac{(2+2\rho-d)\gamma}{2(1- 2\gamma)}} \, \eta(x,\tau).
\end{equation}
This model becomes exactly Eq.~(\ref{kpz2}) in time $\tau$ for $d=2+2\rho$. So choosing appropriate
values of $\rho$ one recovers the KPZ equation in time $\tau$ for any desired spatial dimension $d>2$. Model~(\ref{kpz2})
was analyzed with the SCE and for the dimension under examination $d=2+2\rho$ classical KPZ behavior, as if $\rho=0$, was
found~\cite{ks1}. It is not clear whether or not there exist an upper critical dimension for KPZ
(a dimension above which the large-scale effective behavior of the equation would reduce to that of its linear counterpart)
and what would be its value in the first case~\cite{ps,ks2}. Recent numerical results suggest that,
if it exist, one necessarily has $d_c > 4$~\cite{sp}. In any case, it is clear that for any $d \ge 2$ under the upper critical dimension of KPZ the corresponding dynamic exponent
of model~(\ref{kpz2}) is $z<2$, while the decorrelation threshold is as before anticipated and results $\gamma_d=1/2$. This proves that the nontrivial
coupling of the dilation transformation and the nonlinear field theory extends from two to higher dimensions, at least in the range
$2 \le d \le 4$ according to~\cite{sp}, and possibly to higher dimensions. This also shows that
apparently the one-dimensional situation is left alone as the only one in which the decorrelation threshold is the intuitive one.
And thus, this fact is yet another proof of the fundamentally different character of the KPZ equation in and above one dimension, posing,
in the latter case, a problem much more involved and changeling. This difference should be even more pronounced in the neighboring field of
radial growth, since posing the KPZ problem in this context implies nontrivial topological effects only if $d \ge 2$~\cite{escuderocsf,escudero2}.

As mentioned in the introduction, the condition $\gamma > 1/z$ implies correlations stop propagating in the linear case,
but it is just a necessary and not sufficient condition to achieve the spatial homogeneity of the field. Homogeneity is only
achieved in the large scale if $\gamma > \max \{1/z,1/d\}$ in the linear case and for $\rho=0$~\cite{escuderocsf},
showing that the spatial dimensionality of the system has an important role in this question.
In the linear case and for $\rho>0$ the relation becomes $\gamma > \max \{1/z,1/(d-2\rho)\}$ if $\rho < d/2$; if $\rho \ge d/2$ the
spatial homogeneity of the field is never achieved because this inequality implies correlations do not decay with
distance (a case that is not going to be considered in the following).
But it is not correct employing these relations in the nonlinear setting.
However, it is possible to find an analogous condition by means of the introduction of new critical exponents $\tilde{\alpha}$ and $\tilde{z}$.
To this end we make explicit use of correlations~(\ref{corr2})
and~(\ref{corrd}) for the KPZ case. If we write these correlations in the form suggested by Eq.~(\ref{superprin}), so that the dependence
on the dilation of space becomes explicit, we find the expression
\begin{equation}\label{newexps}
G_{\{2,d\}}^2=t^{2\tilde{\alpha}\gamma}|x-x'|^{2 \tilde{\alpha}}
\left\{\tilde{\mathcal{F}},\tilde{\mathcal{G}} \right\} \left( \frac{t^\gamma|x-x'|}{t^{1/\tilde{z}}} \right),
\end{equation}
where the new exponent values are $\tilde{\alpha}=(1-2\gamma)\alpha_{\mathrm{KPZ}}/(1-2\gamma+z_{\mathrm{KPZ}}\gamma)$
and $\tilde{z}=z_{\mathrm{KPZ}}/(1-2\gamma+z_{\mathrm{KPZ}}\gamma)$ whenever
$\gamma <1/2$, where $\alpha_{\mathrm{KPZ}}$ and $z_{\mathrm{KPZ}}$ are the corresponding
exponents of KPZ at the lower critical dimension $d=2+2\rho$ (including the case $\rho=0$).
For $\gamma \ge 1/2$ the exponents become $\tilde{\alpha}=0$ and $\tilde{z}=2$,
with marginal logarithmic corrections for $\gamma=1/2$. Also,
if one uses the form of both correlations $G=t^{2\beta} \, \mathcal{H}(|x-x'|/t^{1/z})$
one may extract in our case the value of the new growth exponent $\tilde{\beta}=(1-2\gamma)\beta_{\mathrm{KPZ}}$
for $\gamma<1/2$ and $\tilde{\beta}=0$ for $\gamma \ge 1/2$
(with again a marginal logarithmic correction for $\gamma=1/2$)
and find the relation $\tilde{\beta}=\tilde{\alpha}/\tilde{z}$ holds in this case too.
Additionally one sees that using the relation $\alpha_{\mathrm{KPZ}}+z_{\mathrm{KPZ}}=2$
one finds $\tilde{\alpha}+\tilde{z}=2$.
The different relation found for the KPZ equation with a time dependent coefficient of the nonlinearity~\cite{emilio}
can be written in the present terms as $\alpha_{\mathrm{KPZ}}+z_{\mathrm{eff}}=2+2 \gamma z_{\mathrm{eff}}$.
From these results one
reads that the dilation of space systematically neglects the non-perturbational behavior of KPZ.
Another quantity
within reach is the center of mass fluctuations, which can be characterized by a new exponent $\sigma \left(\int \phi (x,t) dx \right) \sim t^\mu$,
where $\sigma(\cdot)$ denotes the standard deviation of the corresponding random variable.
In the present case this exponent reads
$\mu= [\beta_{\mathrm{KPZ}} +(d-2\rho)/(2z_{\mathrm{KPZ}})](1-2\gamma)$ for $\gamma <1/2$ and $\mu=0$ for $\gamma \ge 1/2$, with as previously a logarithmic
correction for $\gamma=1/2$. In particular, the motion of the center of mass becomes bounded in time for $\gamma>1/2$. Note that
this quantity is closely related to the properties characterizing weak convergence of the profile of $\phi$ to the homogeneous
spatial state~\cite{escuderocsf}. In general the exponents which characterize weak convergence to the homogeneous profile are different
from the exponents appearing in the field difference correlation function~(\ref{corrd})~\cite{escuderocsf}. However, in the present cases,
both sets of exponents are exactly the same. This should be considered by no means a general feature of the KPZ equation: it is
a direct consequence of the fact that we are always considering this equation at its lower critical dimension. Note also that our results
imply the flatness of the field (in the sense that both field difference and two-point
correlation functions are uniformly bounded in both space and time)
is achieved for $\gamma > 1/2$. This is in disagreement with the linear requirement $\gamma > \max \{1/z_{\mathrm{KPZ}},1/(d-2\rho)\}$. However,
it is in perfect agreement with the modified requirement $\gamma > \max \{1/\tilde{z},1/(d-2\rho)\}$. In the same way,
the threshold for the loss of correlation can be expressed by the inequality $\gamma > 1/\tilde{z}$. Both inequalities express
the fact that the linear conditions for decorrelation and homogeneity of the field can be extended to the nonlinear case
provided we introduce $\gamma-$dependent exponents. This is another way of expressing that the coupling of the spatial dilation transformation
and the nonlinear dynamics of the field is nontrivial.

\begin{figure}
\centering \subfigure[]{
\includegraphics[width=0.4\textwidth]{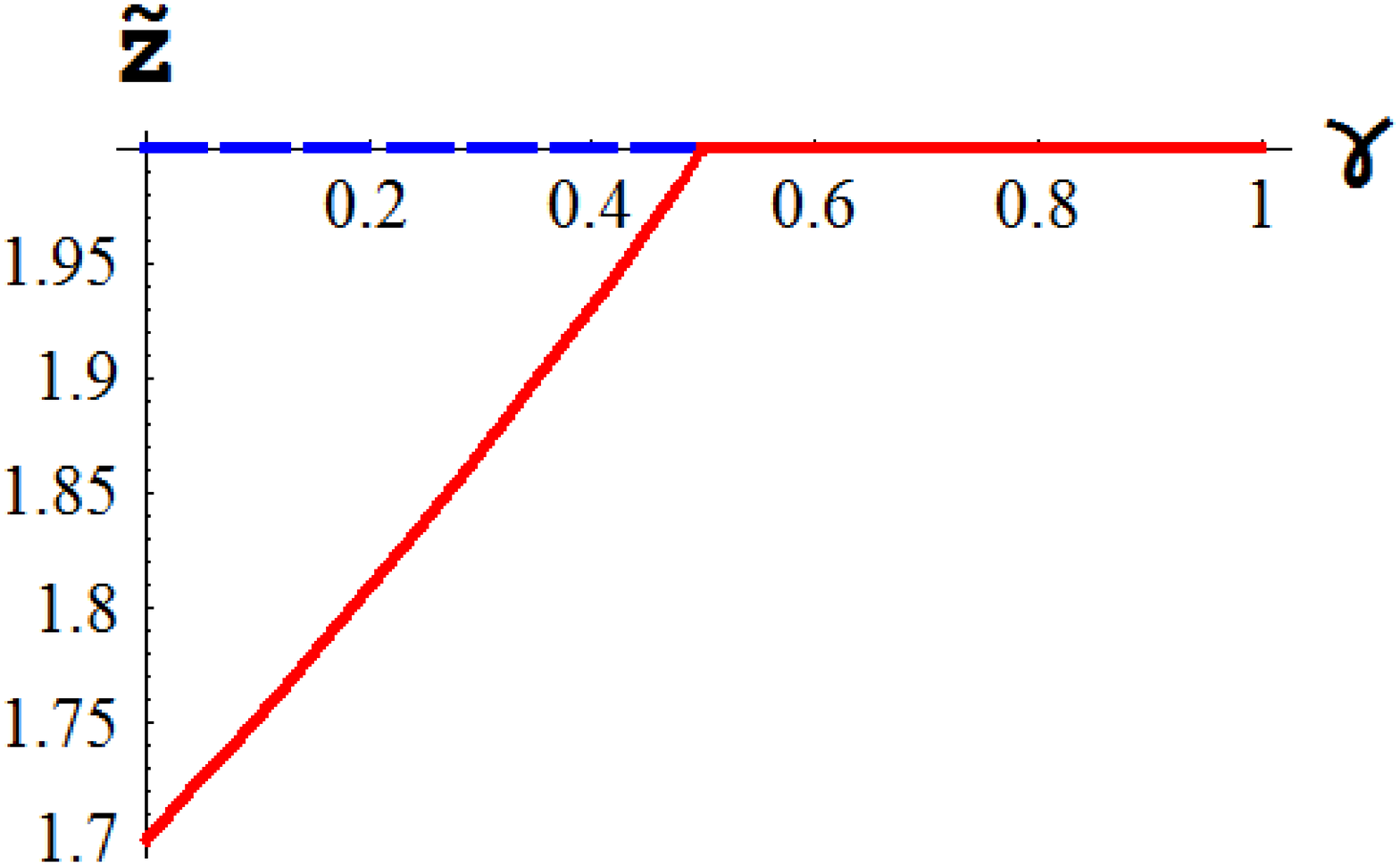}
\label{zg}} \subfigure[]{
\includegraphics[width=0.4\textwidth]{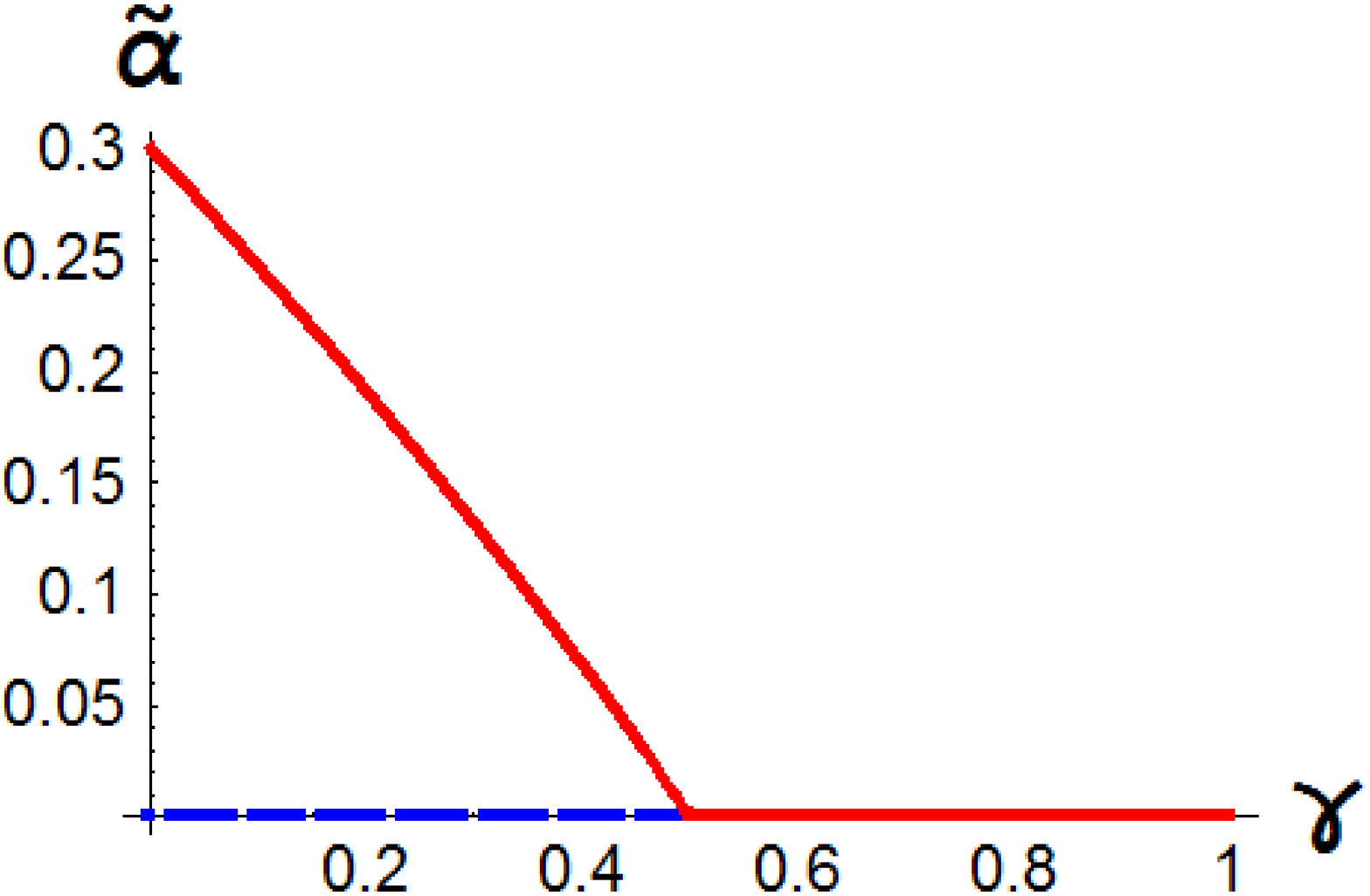}
\label{ag}} \\ \subfigure[]{
\includegraphics[width=0.4\textwidth]{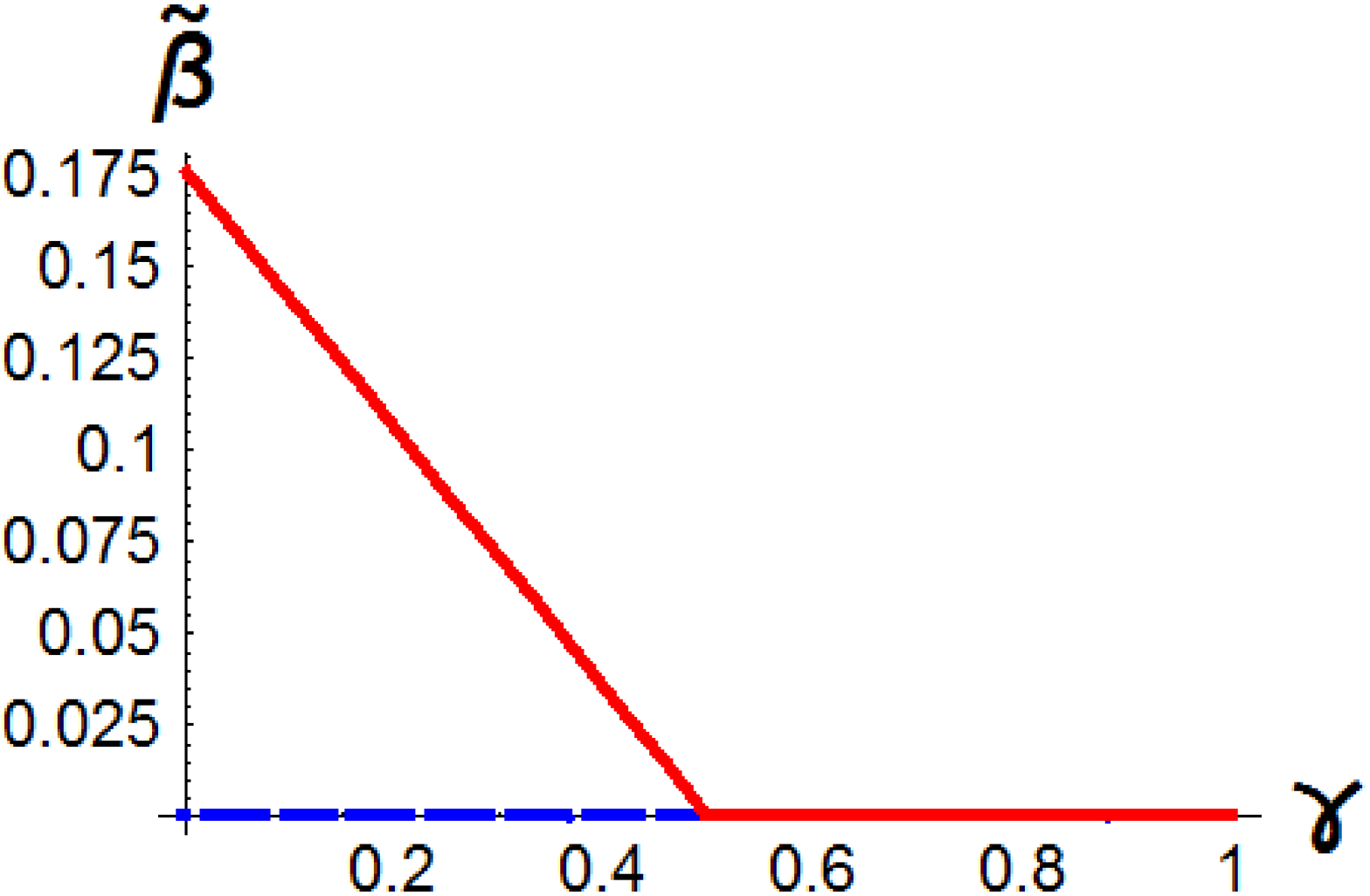}
\label{bg}} \subfigure[]{
\includegraphics[width=0.4\textwidth]{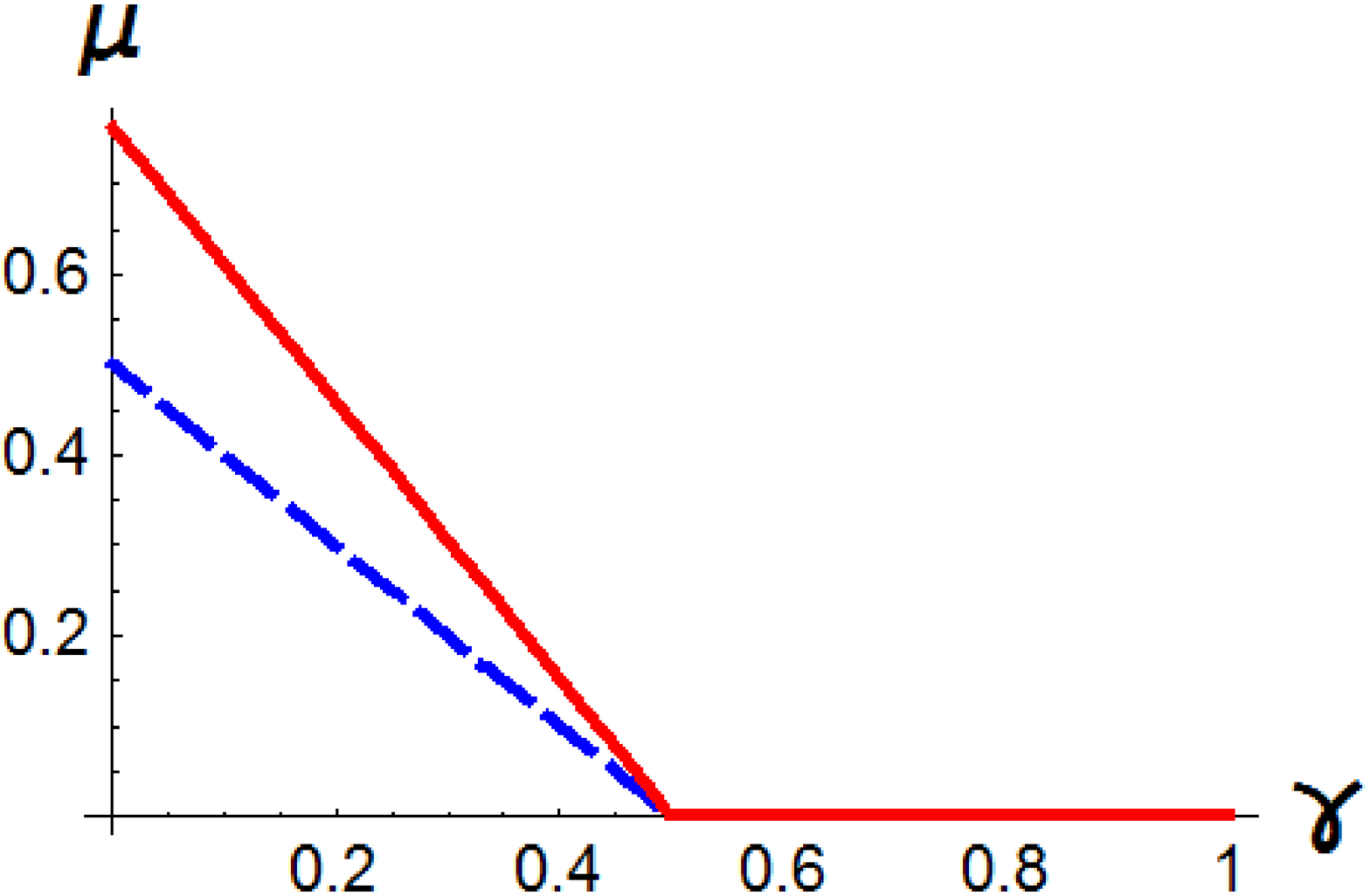}
\label{mg} }
\caption{Exponents $\tilde{z}$, $\tilde{\alpha}$, $\tilde{\beta}$ and $\mu$ versus $\gamma$ for the KPZ (red solid line) and EW (blue dashed line)
equations as explained in the text.}
\label{graph}
\end{figure}

\begin{figure}
\centering \subfigure[]{
\includegraphics[width=0.4\textwidth]{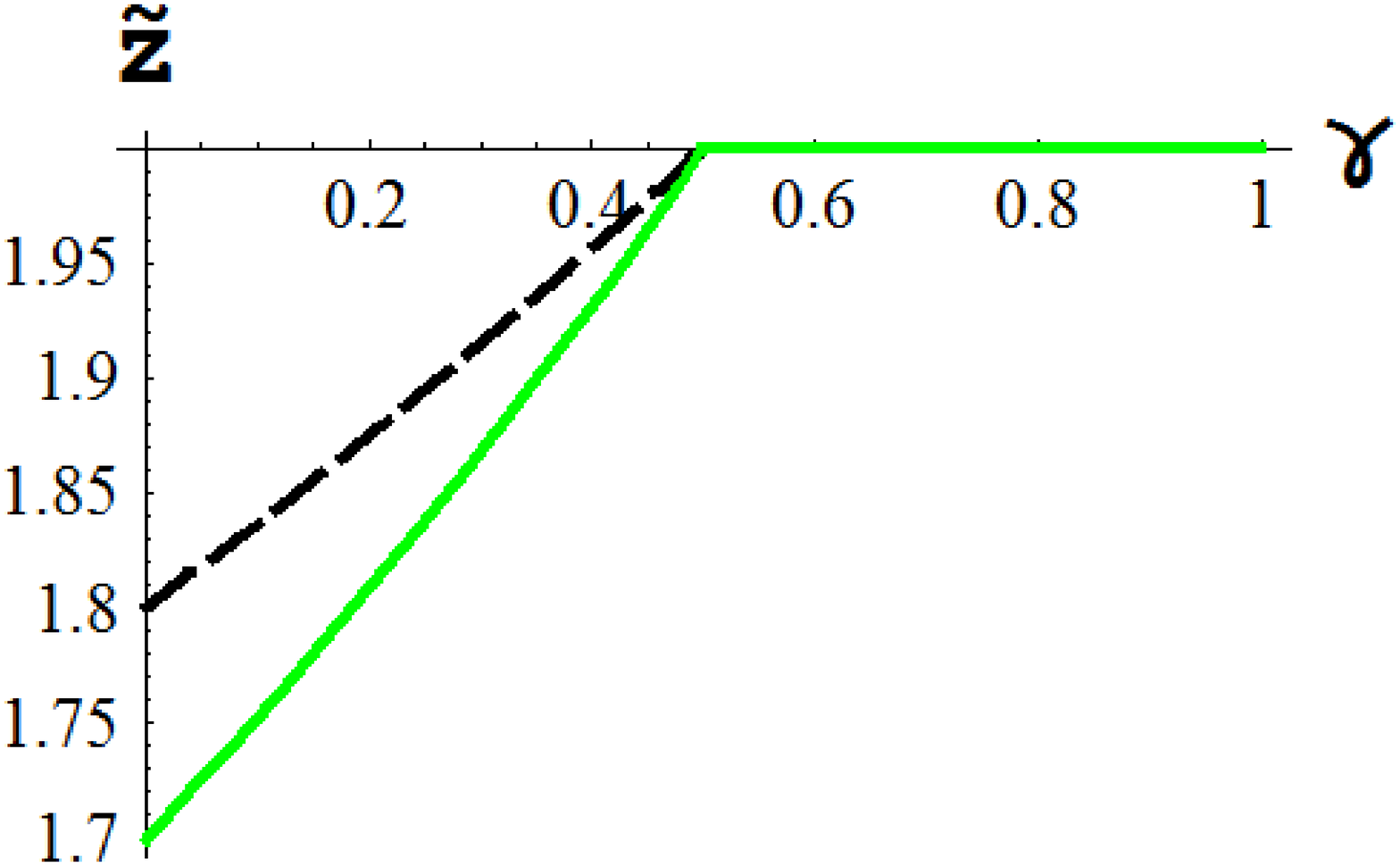}
\label{zg}} \subfigure[]{
\includegraphics[width=0.4\textwidth]{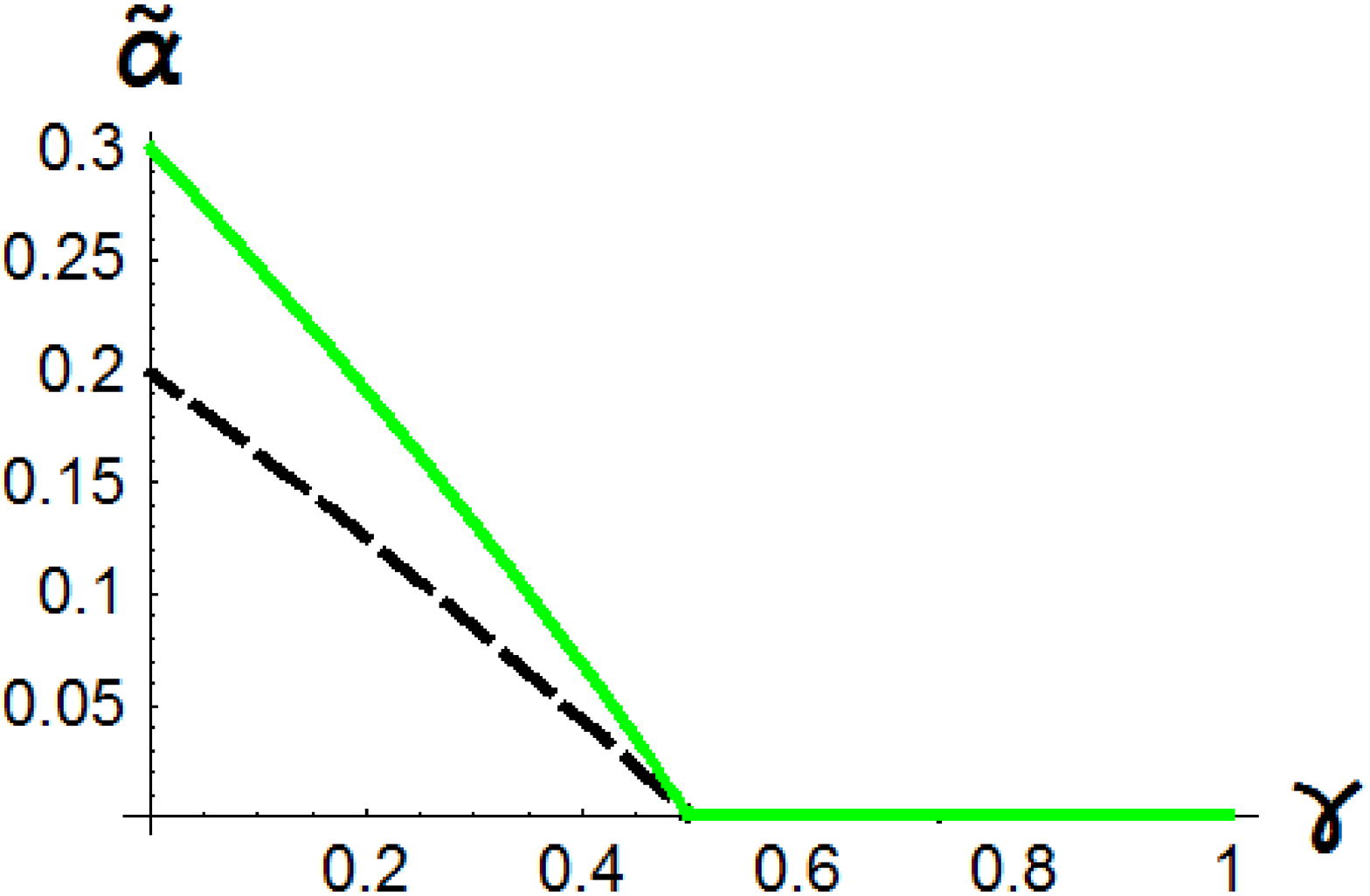}
\label{ag}} \\ \subfigure[]{
\includegraphics[width=0.4\textwidth]{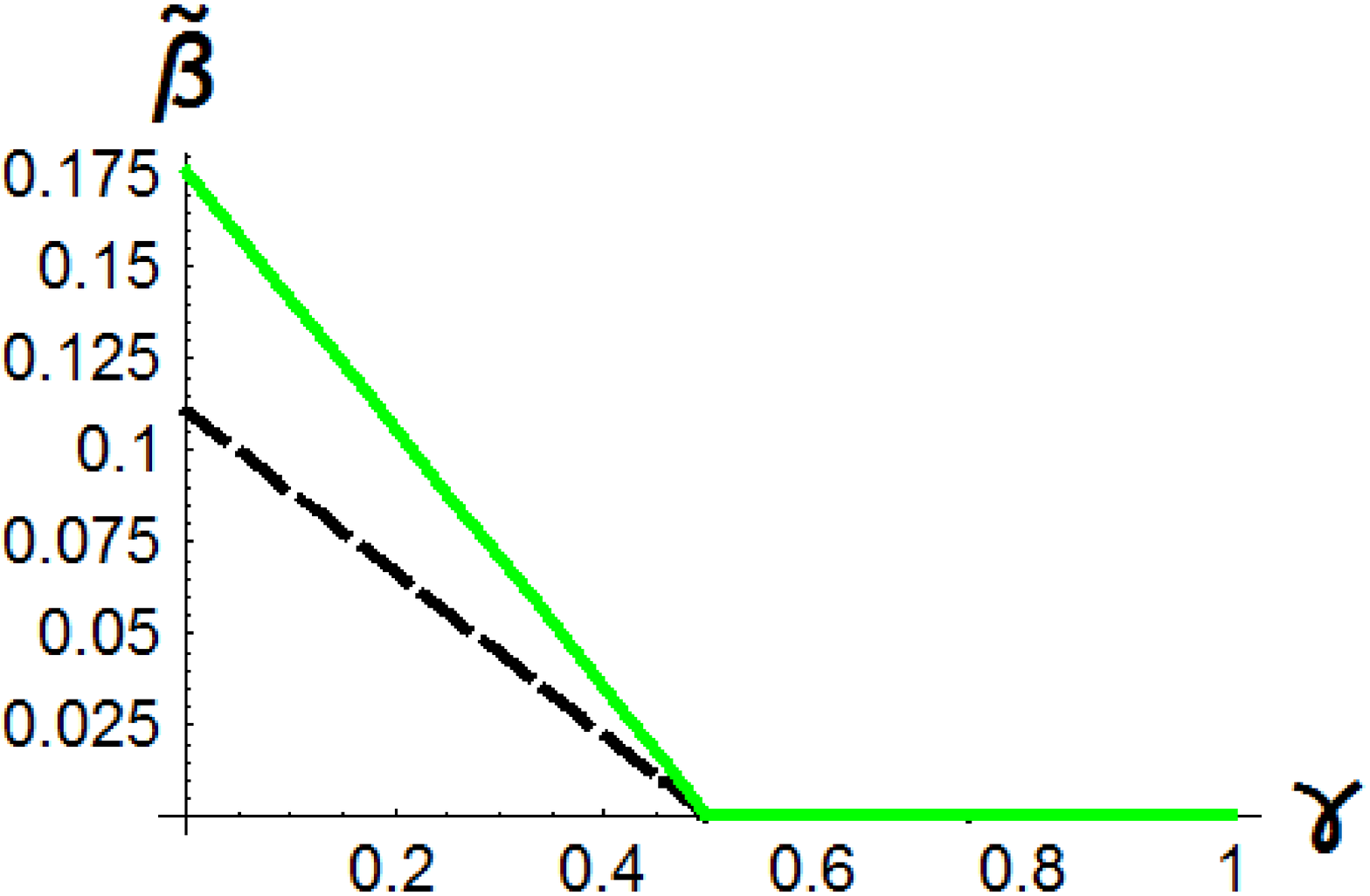}
\label{bg}} \subfigure[]{
\includegraphics[width=0.4\textwidth]{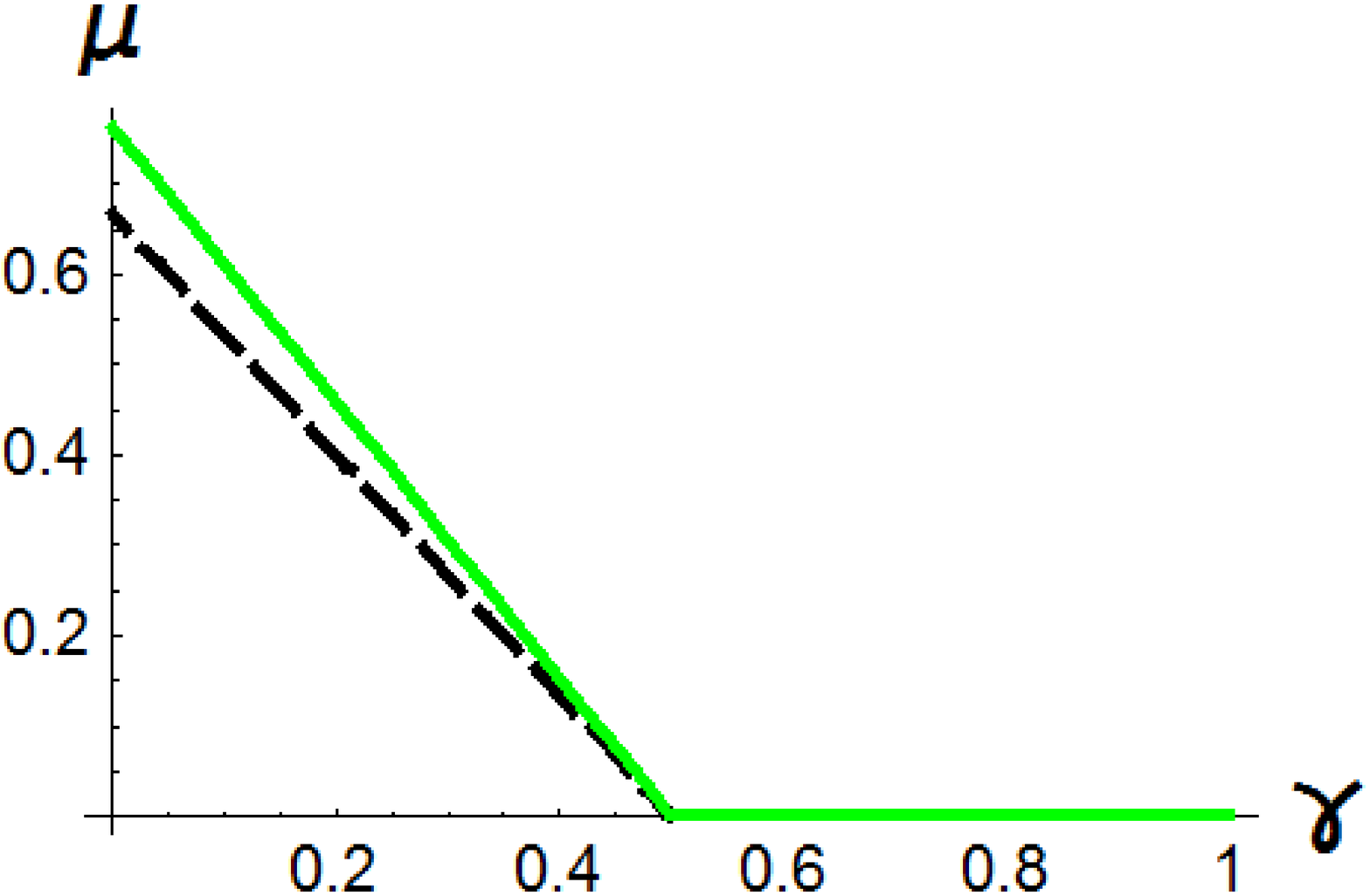}
\label{mg} }
\caption{Exponents $\tilde{z}$, $\tilde{\alpha}$, $\tilde{\beta}$ and $\mu$ versus $\gamma$ for the KPZ equation at $d=2$ (green solid line)
and $d=4$ (black dashed line) as explained in the text.}
\label{graph2}
\end{figure}

We summarize part of our results in Figs.~\ref{graph} and~\ref{graph2}. The exponents $\tilde{z}$, $\tilde{\alpha}$, $\tilde{\beta}$ and $\mu$ corresponding
to Eq.~(\ref{kpz}) are shown in Fig.~\ref{graph} for a varying $\gamma$. Together with them, we show the resulting exponents for the Edwards-Wilkinson (EW)
equation, which is the linearization of the KPZ equation obtained by setting $\lambda=0$ in~(\ref{kpz}). We remark again that the exponents
$\tilde{z}$, $\tilde{\alpha}$ and $\tilde{\beta}$ describe the self-similar behavior of both correlation functions $G_d$ and $G_2$ for both the KPZ
and EW equations. This result is not general but a consequence of the fact that we are considering both equations in $d=2$, that turns out to be
the lower critical dimension of KPZ and the critical dimension of EW. One immediately reads from this figure that exponents
$\tilde{z}$, $\tilde{\alpha}$ and $\tilde{\beta}$ for the EW equation are independent of $\gamma$. On the other hand, these same exponents for the
KPZ equation depend monotonically on $\gamma$ for $\gamma < 1/2$ and become constant and equal to those of the EW equation for $\gamma \ge 1/2$.
The behavior of the exponent $\mu$ is different. It depends monotonically on $\gamma$ for $\gamma < 1/2$ for both the KPZ and EW equation. In this
regime the values of this exponent in the linear and nonlinear cases are different. However, both values become constant and equal for $\gamma \ge 1/2$.
All these results show that the homogeneous spatial dilation effectively linearizes the KPZ equation for $\gamma>1/2$ (we remind that there exist
marginal logarithmic corrections at $\gamma=1/2$) when considered at its lower critical dimension.

We summarize the effect of a non-vanishing $\rho$ in Fig.~\ref{graph2}. The exponents $\tilde{z}$, $\tilde{\alpha}$, $\tilde{\beta}$ and $\mu$
corresponding to Eq.~(\ref{kpz}) (that is, the case $\rho=0$) at $d=2$ and Eq.~(\ref{kpz2}) at $d=4$ (for the corresponding $\rho=1$) are
represented for a varying $\gamma$. From these figures one sees that the results for higher dimensions interpolate between the two-dimensional result
and the EW one. This is not surprising because as we approach the critical dimension of KPZ (be it finite or infinite) the results should be
closer to those of the EW equation. Note that the decorrelation threshold is always $\gamma_d=1/2$. This fact remains true for any $d$ and any $\rho$
as long as we consider Eq.~(\ref{kpz2}) at its lower critical dimension. This result follows immediately from the analytical formulas we have derived
herein, and in particular we see that $\tilde{\beta}/\beta_{\mathrm{KPZ}}=1-2\gamma$ and
$\mu(\gamma)/\mu(0)=1-2\gamma$ for $\gamma \le 1/2$.

We also note that, although we have proven the threshold $\gamma_d=1/2$ for decorrelation and homogeneity of the field
for the lower critical dimension of KPZ, we expect it will stay the same for dimensions above
this one. This conjecture comes from the fact that change of variables~(\ref{change}) sends the equations under consideration to
KPZ equations with noises whose amplitudes depend on a negative power of time in the case of a higher dimensionality.
The question of \emph{super-roughness} of the field, i.~e. finding the values of $\gamma$ for which the fluctuations of the field
grow faster than the dilation of space, is a simple corollary of our results. This would happen whenever $\tilde{\alpha}>1$, what is
impossible for any $\gamma \ge 0$. Finally, we mention that our results are not particular to the KPZ equation, but can be applied to any
nonlinear stochastic partial differential equation of this sort. For instance, let us consider the conserved counterparts of the KPZ equation
\begin{equation}\label{ckpz}
\partial_t \phi= -\nu \nabla^4 \phi + \frac{\lambda}{2} \nabla^2 (\nabla \phi)^2 + \xi^{\{n,c\}}(x,t).
\end{equation}
where the noises $\xi^{\{n,c\}}$ are zero-centered white Gaussian fields whose correlations are respectively given by
\begin{eqnarray}
\left\langle \xi^{n}(x,t) \xi^{n}(x',t') \right\rangle &=& D \delta(x-x') \delta(t-t'), \\
\left\langle \xi^{c}(x,t) \xi^{c}(x',t') \right\rangle &=& -D \nabla^2 \delta(x-x') \delta(t-t').
\end{eqnarray}
It is now straightforward to analyze the effect of a homogeneous spatial dilation on these equations by repeating the arguments
invoked for the KPZ equation. In particular we find for both equations $\gamma_d=1/4$ for $d=4$ (in the case of $\xi^n$) and
for $d=2$ (in the case of $\xi^c$).

In summary, we have studied the effect of a uniform dilation of space on the dynamics
of nonlinear fields theories. In particular we have focused on the nonlinear KPZ equation with different stochastic forcing terms, because this
field theory is known to display nontrivial effects regarding the velocity at which correlations propagate. We have argued that in one dimension
numerical results suggest that the loss of correlation starts when the velocity at which the space grows overtakes the velocity at which correlations
propagate in the absence of spatial dilation. However, in two and higher dimensions the threshold for the appearance of decorrelation becomes anticipated,
and so loss of correlation starts at a velocity of the dilation transformation slower than the speed at which correlations propagate. This fact is
a consequence of the nontrivial behavior of the KPZ equation at its lower critical dimension. It shows that the interplay of spatial dilation and nonlinearity
is far from trivial and, in particular, that it is not possible to infer the effect of a dilation of space on a nonlinear field theory \emph{a priori}.

There are several interesting connections among models in condensed matter physics and cosmology. In this work we have discussed one such model
given by the KPZ equation, which lies in the mentioned interface as well as Ginzburg-Landau theories~\cite{devega} and
Bose-Einstein condensation~\cite{visser}. Some questions naturally emerge from the present study.
One is determining under which conditions
loss of correlation in an anisotropically expanding system is achieved. Mathematically,
accounting for anisotropic expansions implies the substitution of the FLRW
metric by a Bianchi I metric~\cite{cembranos}. Another problem is the analysis of related nonlinear models
with a source of quantum fluctuations instead of the classical ones. In this framework the question of under
which conditions disentanglement occurs~\cite{nambu} seems to be connected with the present discussion.
In the field of condensed matter, a possible physical realization of our results could perhaps be achieved in
experiments of combustion fronts in paper. The KPZ equation has been shown to be able to describe these fronts,
although the measured noise is not necessarily one of those we have considered~\cite{myllys1,myllys2}.
We have shown that changing the noise in Eq.~(\ref{ckpz}) changes the critical dimension, but not the
decorrelation threshold. It would be interesting to know if this result holds in the case of combustion fronts too.

\emph{Acknowledgments:} The author is grateful to Jos\'{e} Cembranos and Emilio Hern\'{a}ndez-Garc\'{\i}a for discussions.
This work has been partially supported by projects MTM2010-18128, RYC-2011-09025 and SEV-2011-0087.


\begin{thebibliography} {99}
\bibitem{foot1} The exponent $\alpha$ might depend on the scale $|x-x'|$ if the variations of the field are different in different scales.
Herein we will consider a constant $\alpha$ because we will always focus on the largest spatial scale.
\end{thebibliography}
\end{document}